\documentclass[aps,pre,groupedaddress]{revtex4-2}
\usepackage{natbib}
\usepackage{amsmath}
\usepackage{epsfig}
\usepackage{color}

\usepackage{graphicx}
\usepackage{amsfonts}
\usepackage{dcolumn}
\usepackage{bm}
\usepackage{color}
\usepackage{hyperref}

\begin{document}
	\bibliographystyle{unsrt}

\title{Computing Large Deviations of First-Passage-Time Statistics in Open Quantum Systems: Two Methods} 

\author{Fei Liu\textsuperscript{1}}
\email{feiliu@buaa.edu.cn}
\author{Jiayin Gu\textsuperscript{2}} \email{gujiayin@njnu.edu.cn}
\affiliation{\textsuperscript{\rm 1}School of Physics, Beihang University, Beijing 100083, China}
\affiliation{\textsuperscript{\rm 2}School of Physics and Technology, Nanjing Normal University, Nanjing 210023, China}
 
\date{\today}

\begin{abstract}{
We propose two methods for computing the large deviations of the first-passage-time statistics in general open quantum systems. The first method determines the region of convergence of the joint Laplace transform and the $z$-transform of the first-passage time distribution by solving an equation of poles with respect to the $z$-transform parameter. The scaled cumulant generating function, which is the logarithm of the boundary values within this region, is subsequently obtained. The theoretical basis is that the dynamics of open quantum systems can be unraveled into a piecewise deterministic process and that a tilted Liouville master equation exists in Hilbert space. The second method uses a simulation-based approach that is built on the wave function cloning algorithm. To validate both methods, we derive analytical expressions for the scaled cumulant generating functions in field-driven two-level and three-level systems. In addition, we present numerical results and cloning simulations for a field-driven system comprising two interacting two-level atoms.
}	  
  
\end{abstract}

\maketitle

\section{Introduction} 
\label{section1}

Large deviations in counting statistics characterize the asymptotic characteristics of the distributions of time-extensive counting variables as the duration of stochastic processes approaches infinity~\cite{Touchette2008,Levitov1996,Bagrets2003,Esposito2009,Garrahan2010,Landi2024}. Meanwhile, large deviations in first-passage-time (FPT) statistics concern with the asymptotic features of the distributions of the time variable when the counting variables first reach specific values and these values tend towards positive or negative infinity~\cite{Budini2014,Saito2016,Gingrich2017,Garrahan2017,Ptaszy2018,Rudge2019,Liu2024,Menczel2026}. Intuitively, these two types of large deviations appear to be interconnected. In fact, Saito and Dhar~\cite{Saito2016}, as well as Gingrich and Horowitz~\cite{Gingrich2017}, demonstrated that for currents in classical stochastic processes, such as the entropy production rate in stochastic thermodynamics~\cite{Seifert2011}, the scaled cumulant generating functions (SCGFs) of these two statistics are inverse functions of each other. In this paper, we term a counting variable ``current-like'' if its value can be positive or negative. In classical stochastic systems and open quantum systems, Budini {\it et al.}~\cite{Budini2014} and Garrahan~\cite{Garrahan2017} reached the same conclusion for another type of counting variables. These variables have nonnegative values and do not decrease over time, for example, the total number of random jumps in stochastic trajectories or dynamic activity~\cite{Garrahan2010}. We refer to them as ``simple counting variables'' to distinguish them from the current-like variables. Very recently, the inverse function relationship has been further shown to hold for current-like counting variables in open quantum systems~\cite{Liu2024,Menczel2026}. Intriguingly, this conclusion is independent of whether quantum systems have unique or multiple steady states~\cite{Menczel2026}. The inverse function relationship is significant in the studies of various uncertainty relations in FPT statistics~\cite{Garrahan2017,Gingrich2017,Hiura2021,Pal2021a,Hasegawa2022,Yin2025,Neri2025}. 

The inverse function relationship provides an approach for computing the SCGFs in FPT statistics. First, we obtain the SCGFs in the counting statistics. This process is equivalent to finding the largest real part of the eigenvalues of the generator of the tilted classical master equation~\cite{Levitov1996,Lebowitz1999,Andrieux2008} or the quantum master equation~\cite{Mollow1975,Bagrets2003,Esposito2009,Liu2016a}. Second, we perform an inverse function transform (swapping the positions of the variable and the function in the data). Despite this convenience, it is interesting to investigate whether alternative methods exist for computing the large deviations in FPT statistics. FPT statistics is related to, but not equivalent to, counting statistics ~\cite{Rudge2019,Menczel2026,Ptaszy2018}. Thus, developing its own theories or methods is worthwhile. On the other hand, computing large deviations in counting statistics is challenging~\cite{Giardina2006,Lecomte2007,Cavallaro2016}. However, prior to a comprehensive theoretical exploration, we cannot exclude the possibility that computing the large deviations in the FPT statistics might be easier. If this is truly the case, because the inverse function relationship is bidirectional, then it is equivalent to presenting a new approach for computing the difficult large deviations in the counting statistics. This phenomenon was indeed witnessed in resonantly driving two-level quantum systems~\cite{Liu2024}, where several analytical SCGFs were obtained in the FPT statistics. Our motivation becomes more obvious when we imagine a scenario in which the inverse function relationship has not been discovered, yet we still aim to evaluate the large deviations in the FPT statistics.  

In this paper, we propose two methods for computing the large deviations in FPT statistics of general open quantum systems. The first method solves an equation of poles. This extends the method previously proposed for a specific class of open quantum systems~\cite{Liu2024}, where quantum jump trajectories can be interpreted as a semi-Markov process~\cite{Agarwal1977,Carollo2019,Liu2022}. We emphasize that this effort is significant, because the open quantum systems described by semi-Markov processes are atypical; they are exceptions~\cite{Hegerfeldt1993,Soler2008,Landi2024}. To derive the equation of poles, we use the piecewise deterministic processes of the wave function~\cite{Breuer2002} and the tilted Liouville master equation in Hilbert space~\cite{Liu2021a}, and analyze the region of convergence of the joint Laplace transform and the $z$-transform of the FPT distribution~\cite{Liu2024}. An advantage of this method is its unified treatment of both types of counting variables. The second method is a simulation-based approach that is built upon the wave function cloning algorithm. This algorithm is a variant of the one used in counting statistics~\cite{Carollo2019}. When open quantum systems have a large number of degrees of freedom, simulations become essential. In these cases, numerically solving the equation of poles becomes inefficient because of the large size of the associated matrix.

The remainder of this paper is organized as follows. In Sec.~(\ref{section2}), we review the counting statistics in the Hilbert space. In addition to defining relevant terms and notations, we present the joint transition probability functional and provide its formal solution. On the basis of these results, in Sec.~(\ref{section3}), we extend the formula proposed by Saito and Dhar to the Hilbert space of general open quantum systems. In Sec.~(\ref{section4}), we derive the large deviations in FPT statistics and their connections to the large deviations in counting statistics, all within the Hilbert space. To make the method practically feasible, we present an equation of poles in averaged dynamics. In Sec.~(\ref{section5}), we develop a simulation-based approach built on a cloning algorithm. In Sec.~(\ref{section6}), we illustrate the two methods using field-driven two-level and three-level systems, and a system composed of two interacting two-level atoms. Section~(\ref{section7}) concludes this work. The reader is reminded that most of the derivations presented in this paper are formal; mathematical rigor is not the primary objective of this paper.  

\section{Counting statistics in Hilbert space}  
\label{section2}
 
\subsection{Piecewise deterministic processes} 
\label{section21}
We begin with the reduced density matrix $\rho(t)$ of an open quantum system. Under appropriate assumptions and conditions, its dynamics is described by the quantum master equation~\cite{Davies1974,Lindblad1976,Gorini1976} 
\begin{eqnarray}
	\label{quantummasterequation}
	\partial_t \rho(t)&=&-\text{i}[H,\rho(t)]+\sum_{\alpha=1}^M r_\alpha\left( A_\alpha\rho(t){A_\alpha}^\dag -\frac{1}{2}\left\{{A_\alpha}^\dag A_\alpha,\rho(t)\right\}\right)\\
	&\equiv & L \rho(t),
\end{eqnarray}
where the Planck constant $\hbar$ is set to 1, $H$ represents the Hamiltonian of the system, and $A_\alpha$ and $r_\alpha$, for $\alpha=1,\cdots,M$, represent the jump operators and positive rates, respectively. For simplicity, we assume the system has a unique steady sate.  Equation~(\ref{quantummasterequation}) can be unraveled into quantum jump trajectories~\cite{Carmichael1989,Dalibard1992,Dum1992,Wiseman1993}. These trajectories are composed of alternating deterministic pieces and random jumps of the wave functions of individual quantum systems. The basic stochastic process is the piecewise deterministic process of the wave function~\cite{Breuer2000}. If the probability distribution functional of the wave function $\psi$ at time $t$ is ${\cal P}[\psi|t]$, the reduced density matrix is equal to the ensemble average of the pure states, {\it i.e.}, 
\begin{eqnarray}
{\rho}(t)=\int D\psi D\psi^*{\cal P} [\psi|t] |\psi\rangle\langle \psi|, 
\end{eqnarray}
where $D\psi D\psi^*$ represents the Hilbert space volume element. In this regard, the quantum master equation~(\ref{quantummasterequation}) describes the averaged dynamics. The functional ${\cal P}[\psi|t]$ has been shown to satisfy the Liouville-master equation of the piecewise deterministic process in the Hilbert space~\cite{Breuer1997a,Breuer2000,Breuer2002}. Because this equation is not be used in this paper, we do not explain it further. Instead, we make only a brief comment when we review the counting statistics below. 

\subsection{Tilted Liouville master equation}
\label{section22}
The quantum jump trajectory forms the basis of the counting statistics~\cite{Srinivas1981,Carmichael1989,Levitov1996,Plenio1998,Bagrets2003,Brandes2008,Esposito2009,Garrahan2010,Bruderer2014,Landi2024}. Consider a trajectory with $N$ jumps. Let the instant and type of each jump be $t_i$ and $\alpha_i$, respectively, for $i=1,\cdots, N$. We denote this trajectory as   
\begin{eqnarray}
\label{trajectorydefinition}
	{X}_N(t)=(\alpha_{N},\cdots, {\alpha_2},{\alpha_1}),
\end{eqnarray}
where $t$ represents the duration of the process. We have ordered these jumps from right to left in chronological sequence. The jump type $\alpha$ is specified by the jump operator $A_\alpha$ corresponding to each jump event. Thus, $\alpha$ can take on one of the values of $1,\cdots,M$. In Eq.~(\ref{trajectorydefinition}), we define the counting variable as follows: 
\begin{eqnarray}
\label{countingquantity}
	C[{X}_N(t)]=\sum_{i=1}^N w_{\alpha_i},
\end{eqnarray}
where the coefficient $w_{\alpha_i}$ represents a weight assigned according to the jump type $\alpha_i$ at the jump instant $t_i$. Owing to the random nature of the quantum jump trajectories, the value of Eq.~(\ref{countingquantity}) is also random. Let the value be $n$. Counting statistics focuses on the computations and properties of the probability distribution $P(n|t)$. Because this paper is centered on FPT statistics, we concentrate on specific weights that are simply equal to $\pm 1$ or $0$. As a result, the values $n$ of the counting variable are always discrete integers, and their temporal changes are $\pm 1$ or 0. As originally mentioned, we classify counting variables as simple or current-like.

Rather than the probability distribution $P(n|t)$, counting statistics generally focuses on its moment generating function. Because the value of the counting variable in Eq.~(\ref{countingquantity}) is discrete, this function is also the $z$-transform of $n$~\cite{Oppenheim1997}:    
\begin{eqnarray}
\label{momentgeneratingfunctionasztransform}
	{\bar P}(z|t)=\sum_{n=-\infty}^{+\infty} P(n|t)z^{-n},
\end{eqnarray}
where the parameter $z$ is restricted to be positive real. In the remainder of this paper, we use ``bar'' above a symbol to denote the $z$-transform of the counting variable $n$. In the Hilbert space, the moment generating function can be rewritten as  
\begin{eqnarray}
\label{momentgeneratingfunctionforcountingstatistics} 
\int D\psi D\psi^* \bar {\cal P}[\psi,z|t],
	\end{eqnarray}
where $\bar {\cal P}[\psi,z|t]$ represents the $z$-transform of ${\cal P}[\psi,n|t]$. The latter, ${\cal P}[\psi,n|t]$, denotes the joint distribution functional in which the wave function of a quantum system is $\psi$, whereas the value of Eq.~(\ref{countingquantity}) is $n$ at time $t$. An advantage of this alternative expression is that the transformed functional satisfies the tilted Liouville master equation~\cite{Liu2021a}: 
\begin{eqnarray}
	\label{tiltedLiouvillmasterequation}
	\partial_t \bar {\cal P} [\psi,z|t]&=&\text{i}\int dx\left ( \frac{\delta}{\delta \psi(x)}G[\psi](x)-\frac{\delta}{\delta \psi^*(x)}G[\psi]^*(x) \right )\bar {\cal P} [\psi,z|t] \nonumber \nonumber \\
	&&+\int D\phi D\phi^* \left ( \bar {\cal P} [\phi,z|t]W_z[\phi|\psi]-\bar {\cal P} [\psi,z|t]W[\psi|\phi] \right )\nonumber \\
	&\equiv&  \tilde{\cal L}_z[\psi]\bar {\cal P}[\psi,z|t], 
\end{eqnarray}
where $\delta/\delta \psi(x)$ and $\delta/\delta^* \psi(x)$ represent functional derivatives and $x$ represents the positional coordinate. The explicit expressions for the operator $G[\psi]$, the transition rates $W[\psi|\phi]$ and its tilted counterpart $W_z[\phi|\psi]$ on the right-hand side of the equation are provided in Appendix~\ref{detailsintiltedLiouvillemasterequation}. If $z=1$, Eq.~(\ref{tiltedLiouvillmasterequation}) reduces to the Liouville-master equation of the piecewise deterministic process. In this case, $\bar{\cal P}[\psi|t]$ is the probability distribution functional ${\cal P}[\psi|t]$ mentioned in Sec.~(\ref{section21}). The two integrals represent the factors causing changes in the distribution: the first integral corresponds to the deterministic dynamics, and the second integral corresponds to the random jumps of the wave function.
 
\subsection{Joint transition probability functional}
\label{section23}
The piecewise deterministic process of the wave function is Markovian~\cite{Davis1984}. Consequently, we can formally define the joint transition probability functional
$T[\psi,n|\varphi,t]$. This functional represents the possibility that the quantum system begins with the wave function $\varphi$ at time $0$, and at time $t$, the wave function and the counting variable in Eq.~(\ref{countingquantity}) are $\psi$ and $n$, respectively. Equation~(\ref{tiltedLiouvillmasterequation}) yields a formal solution, and its $z$-transform is expressed as    
\begin{eqnarray}
\label{ztransformofjointprobabilityfunctional}
{\bar T}[\psi,z|\varphi,t]=e^{\tilde{\cal L}_z[\psi] t}\delta[\psi-\varphi].
\end{eqnarray}
With this result, we derive the Laplace transform of time $t$ for the moment generating function:   
\begin{eqnarray}
\label{momentgeneratingfunctionasjointzandLaplacetransforms}
	 	\hat {\bar P}(z|\nu)=\int D\psi D\psi^\star \frac{1}{\nu-\tilde{\cal L}_z[\psi]}{\cal P}_0[\psi],
\end{eqnarray} 
where ${\cal P}_0[\psi]$ represents the initial distribution functional of the quantum system in the wave function $\psi$. Throughout the remainder of this paper, we use ``hat'' above a symbol to denote the Laplace transform of time $t$. The reason for expressing  Eq.~(\ref{momentgeneratingfunctionasjointzandLaplacetransforms}) in the frequency domain rather than the time domain becomes evident shortly. We emphasize that  Eq.~(\ref{momentgeneratingfunctionasjointzandLaplacetransforms}) is valid only within a region of convergence in the real $\nu$-$z$ plane, where 
\begin{eqnarray}
\label{convergencycondition}
\det[\nu-\tilde{\cal L}_z[\psi]]>0. 
\end{eqnarray}
The reason for this inequality is that the eigenvalues of the operator $\tilde{\cal L}_z[\psi]$ consist of negative real numbers or complex pairs with negative real parts, which is implied in the structure of Eq.~(\ref{tiltedLiouvillmasterequation}). The implications of this region for large deviations are discussed in Sec.~(\ref{section4}).  
 
\section{First passage time statistics in Hilbert space}
\label{section3} 
The FPT statistics focuses on the probability distribution $F(T|n)$. This distribution characterizes the probability that the counting variable in Eq.~(\ref{countingquantity}) first reaches a given threshold value $n$ at time $T$~\cite{Budini2014,Saito2016,Garrahan2017,Gingrich2017,Ptaszy2018,Rudge2019}. Let us define a functional $F[\phi,T|\varphi,n]$, which represents the joint FPT probability distribution. Specifically, given that the quantum system starts with the wave function $\varphi$ at time $0$, this functional describes the probability that the counting variable first reaches the value $n$ at time $T$ and the instantaneous wave function of the system is $\phi$ simultaneously. Clearly, 
\begin{eqnarray} 
	\label{FPTdistribution}
	F(T|n)=\int D\varphi D\varphi^*\int D\phi D\phi^* F[\phi,T|\varphi,n]{\cal P}_0[\varphi].
\end{eqnarray} 

In classical Markov processes, Saito and Dhar derived a probability relationship between the joint FPT probability distribution and the joint transition probability function~\cite{Saito2016}. Because the piecewise deterministic process of the wave function is essentially Markovian, this relationship also holds in the Hilbert space, with only a minor adjustment in notations. We express it in the frequency domain as follows：
\begin{eqnarray}
\label{functionalSatioDharequation}
{\hat T}[\psi, n|\varphi,\nu] =\int D\phi D\phi^* \hat{T}[\psi,0|\phi,\nu]{\hat F}[\phi,\nu|\varphi, n],
\end{eqnarray}
where $\hat T[\psi,n|\varphi,\nu]$ represents the Laplace transform of the joint transition probability functional $T[\psi,n|\varphi,t]$. Because we are interested in FPT statistics, it is more advantageous to solve for $\hat F$ in terms of $\hat T$ using Eq.~(\ref{functionalSatioDharequation}). We assume the existence of an inverse functional ${\hat T}^{-1}$ that satisfies 
\begin{eqnarray}
	\label{inversT}
	\int D\psi D\psi^*\hat{T}^{-1}[\psi,0|\phi',\nu]
	\hat{T}[\psi,0|\phi,\nu]=\delta[\phi'-\phi]. 
\end{eqnarray}
By performing a joint Laplace transform of the time $T$ and $z$-transform of the counting variable $n$ on both sides of Eq.~(\ref{FPTdistribution}), using Eq.~(\ref{inversT}) to solve for $\hat F$ in Eq.~(\ref{functionalSatioDharequation}), and substituting the solution, we establish a connection between the FPT statistics and the stochastic dynamics of the piecewise deterministic process in the Hilbert space: 
\begin{eqnarray}
	\label{jointzandLaplacetransformFPTdistribution}
	\bar{\hat F}(\nu|z)=\int D\phi D\phi^* \int D\psi D\psi^*
	\hat{T}^{-1}[\psi,0|\phi,\nu] \frac{1}{\nu-\tilde{\cal L}_z[\psi]}
	{\cal P}_0[\psi]. 
\end{eqnarray} 
To obtain this result, we assume that $\bar{\hat T}=\hat{\bar T}$. Equation~(\ref{jointzandLaplacetransformFPTdistribution}) clarifies why we used Eq.~(\ref{momentgeneratingfunctionasjointzandLaplacetransforms}) in the frequency domain: Both integral equations share the same inverse operator, $1/(\nu-\tilde{\cal L}_z[\psi])$. Consequently, they share the same region of convergence in the real $\nu$-$z$ plane. This observation indicates that, for the piecewise deterministic process of the wave function in the Hilbert space, there is an intimate relationship between the counting statistics and FPT statistics. 

\section{Large deviations in two statistics and their connections}
\label{section4}
If we disregard the differences in notation and terminology, on the basis of their structures and probability interpretations, Eqs.~(\ref{momentgeneratingfunctionasjointzandLaplacetransforms}) and~(\ref{jointzandLaplacetransformFPTdistribution}) are identical to those obtained in the classic stochastic processes~\cite{Saito2016,Gingrich2017,Ptaszy2018}, for example, Eqs.~(2)-(7) in the supplemental material of Ref.~\cite{Gingrich2017}. This finding is not surprising, as all the derivations are based on Markovian characteristics. Thus, we can directly extend the established results of the large deviations in counting statistics and FPT statistics~\cite{Saito2016,Gingrich2017,Liu2024} to general open quantum systems. Given the detailed explanations of these results in Ref.~\cite{Liu2024}, we only summarize them with a minor adjustment to the notation.  

First, the boundary of the region of convergence is an open curve that passes through the point $(0,1)$ and opens to the right in the real $\nu$-$z$ plane. Significantly, this boundary curve satisfies the equation of poles of Eqs.~(\ref{momentgeneratingfunctionasjointzandLaplacetransforms}) or~(\ref{jointzandLaplacetransformFPTdistribution}): 
\begin{eqnarray}
	\label{equationofpolesHilbertspace}
	\det[\nu-{\cal L}_z[\psi]]=0,
\end{eqnarray}
which is implied in the definition of the region of convergence. All real solutions of Eq.~(\ref{equationofpolesHilbertspace}) appear as curves in the plane. The boundary curve is unique: When it is observed in the $\nu$ direction, this curve lies on the rightmost side of the plane; when it is observed in the $z$ direction, the curve either extends to infinity or is arch-shaped and is located in the innermost part of all curves. These two cases correspond to simple counting variables and current-like counting variables, respectively. This striking difference in the shapes has a simple explanation from the perspective of the FPT statistics: in the latter case, there are two distinct FPT distributions as the current-like counting variables tend to positive and negative infinity, respectively, whereas in the former case, there is only one distribution.  

Second, for a given real $z$-value, the SCGF $\Phi(\lambda)$ in the counting statistics is the $\nu(z)$-value on the boundary of the region of convergence, where $z=\exp(\lambda)>0$. In fact, there is an alternative expression of this function: the largest real part of all roots $\nu_k(z)$, for $k=1,\cdots$, of Eq.~(\ref{equationofpolesHilbertspace}) with respect to $\nu$, that is, 
\begin{eqnarray}
	\label{SCGFformulacountingstatistics}
	\Phi(\lambda)=\max_k\{\text{Re}[\nu_k(z)]\}.
\end{eqnarray} 
Because the boundary curve lies in the real plane, the right-hand side of Eq.~(\ref{SCGFformulacountingstatistics}) can be replaced with the largest among all the real roots of the equation of poles in the variable $\nu$. The term ``root'' appears because the left-hand side of Eq.~(\ref{equationofpolesHilbertspace}) is a polynomial with respect to the variables $\nu$ and $z$, which we will prove below. The reader is also reminded that all $\nu_k(z)$ are precisely the eigenvalues of the operator ${\cal L}_z[\psi]$.
  
Third, the SCGF in the FPT statistics is somewhat more complex. For simple counting variables, such as dynamical activity~\cite{Budini2014,Garrahan2017}, given a real $\nu$-value, this function is the logarithm of the $z(\nu)$-value on the boundary of the region of convergence. Like Eq.~(\ref{SCGFformulacountingstatistics}), this function also has an alternative expression:   
\begin{eqnarray}
	\label{SCGFdefinitionFPTtype1}
	\Psi(\nu)=\ln \max_m\{ |z_m(\nu)| \}, 
\end{eqnarray} 
where $z_m(\nu)$, for $m=1,\cdots$, represent all the roots of the equation of poles~(\ref{equationofpolesHilbertspace}) with respect to the variable $z$. Again, because the boundary curve is in the real plane, the right-hand side of Eq.~(\ref{SCGFdefinitionFPTtype1}) can be replaced by finding the largest among all real roots of the equation of poles in $z$ and then taking its logarithm. For current-like counting variables, given a real $\nu$-value, the SCGFs are the logarithms of the two $z(\nu)$-values on the boundary curve (the arch-shaped curve). Let us denote them by $z_+(\nu)$ and $z_-(\nu)$ with the latter being greater. These functions are  
\begin{eqnarray}
	\label{SCGFdefinitionFPTtype2positive}
	\Psi_\pm(\nu)=\ln z_\pm(\nu),
\end{eqnarray}
respectively. We emphasize that in either case, there is a lower bound in the range of $\nu$, which is determined by the region of convergence.   

Fourth, the SCGFs in the counting statistics and the FPT statistics are inverse functions of each other, that is, $\Psi=\Phi^{-1}$, or $\Psi_{\pm}=\Phi^{-1}$. 
In the latter case, the inverse function $\Phi^{-1}$ is defined with respect to the upper and lower curves or the two branches on the same curve. This is a natural consequence of the previous three results. 

Two comments are in order. One is that the first result provides an approach for determining the region of convergence in the real plane. The other is that the region of convergence can be defined in the real $\nu$-$\lambda$ plane because $z=\exp(\lambda)$. Consequently, the description and characteristics of the region in the real $\nu$-$z$ plane also apply to the region in the real $\nu$-$\lambda$ plane. 

\subsection{Equation of poles in averaged dynamics} 
\label{section41}

Before discussing the simulation method, we encounter a difficulty in solving Eq.~(\ref{equationofpolesHilbertspace}) with respect to $\nu$ or $z$ because the operator $\tilde{\cal L}_z$ is defined in the abstract Hilbert space. However, this challenge can be avoided by constructing an equivalent yet computable equation. Let the eigenvalue and corresponding eigenfunctional of the generator of the tilted Liouville master equation be $\nu_k(z)$ and ${\cal R}_k[\psi,z]$, respectively, that is, 
\begin{eqnarray}
	\label{righteigenvectorequsHilbertspace}
	\tilde{\cal L}_z[\psi]{\cal R}_k[\psi,z]=\nu_k(z){\cal R}_k[\psi,z]. 
\end{eqnarray}
Because the generator is non-Hermitian, a corresponding left eigenfunctional with the same eigenvalue exits. We do not discuss them further in this paper. In the following, we show that 
\begin{eqnarray}
	\label{rightvectorofgeneratoroftiltedquantummasterequation}
	\varrho_k(z) =\int D\psi D\psi^* {\cal R}_k[\psi,z]|\psi\rangle  \langle\psi|,
\end{eqnarray}
for $k=1,\cdots$, are the eigenoperators of the generator ${\tilde L}_z$ of the tilted quantum master equation~\cite{Mollow1975,Bagrets2003,Esposito2009,Liu2016a,Landi2024} with the same eigenvalues, that is,
\begin{eqnarray}
\label{eigenoperatoroftiltedquantumasterequation}
{\tilde L}_z[\varrho_k(z)]
	&\equiv &-\text{i} [H,\varrho_k(z)]
	+\sum_{\alpha=1}^M r_\alpha \left ( z^{-\omega_\alpha} A_\alpha \varrho_k(z) A^\dag_\alpha -\frac{1}{2}\left\{ A^\dag_\alpha A_\alpha,\varrho_k(z) \right\} \right )\nonumber \\
	&=&\nu_k(z){\varrho}_k(z).
\end{eqnarray}
Consequently, the abstract Eq.~(\ref{equationofpolesHilbertspace}) is replaced by 
\begin{eqnarray}
\label{equationofpolestiltedquantummasterequation}
\det [\nu-\tilde{L}_z]=0,
\end{eqnarray}
which can be easily computed in an appropriate representation, such as the conventional energy representation. Obviously, Eq.~(\ref{equationofpolestiltedquantummasterequation}) is an algebraic polynomial involving the two variables $\nu$ and $z$. Intriguingly, many years ago, Bruderer {\it et al.} used the same equation to reconstruct the generator of the quantum master equation compatible with the observed cumulants of the counting variables. They referred to  Equation~(\ref{equationofpolestiltedquantummasterequation}) as the characteristic polynomial of the tilted generator~\cite{Bruderer2014}. 

The equivalence of Eqs.~(\ref{equationofpolesHilbertspace}) and~(\ref{equationofpolestiltedquantummasterequation}) is demonstrated as follows: The eigenvalues $\nu_k(z)$ of the operator $\tilde{\cal L}_z[\psi]$ indicate that 
\begin{eqnarray}
	\tilde{\cal P}_k[\psi,|t]=e^{\nu_k(z)t}{\cal R}_k[\psi,z]
\end{eqnarray}
is a solution to the tilted Liouville master equation~(\ref{tiltedLiouvillmasterequation}). In accordance with a previously established conclusion~\cite{Liu2021a}, the operator 
\begin{eqnarray}
	\tilde {\rho}_k(z,t)=\int D\psi D\psi^*\tilde{\cal P}_k[\psi,|t] |\psi\rangle \langle \psi|
\end{eqnarray}
satisfies the tilted quantum master equation, 
\begin{eqnarray}
	\label{atitledquantummasterequation}
	\partial_t 	\tilde {\rho}_k(z,t)=\tilde{L}_z[	\tilde {\rho}_k(z,t)]. 
\end{eqnarray}
On the other hand, owing to the definition in Eq.~(\ref{rightvectorofgeneratoroftiltedquantummasterequation}),  $\tilde{\rho}_k(z,t)=\exp[\nu_k(z)t]\varrho_k(z)$. Substituting this equation into Eq.~(\ref{atitledquantummasterequation}), we immediately obtain Eq.~(\ref{eigenoperatoroftiltedquantumasterequation}). 

In summary, we have presented a comprehensive method for computing the large deviations in the FPT statistics. This method involves finding the roots of the equation of poles~(\ref{equationofpolestiltedquantummasterequation}) in the variable $z$ and applying the third result. It does not rely on the large deviations of the counting statistics and the inverse function relationship, that is, the fourth result. 
 
\section{ Wave function cloning algorithm}
\label{section5}
    
When an open quantum system with a large Hilbert space is investigated, determining the large deviations in the FPT statistics by finding the roots of Eq.~(\ref{equationofpolestiltedquantummasterequation}) with respect to the $z$ variable becomes infeasible. This is frequently the case for many-body systems, where the Hilbert space grows exponentially with the degrees of freedom. The same problem also arises in computing large deviations in the counting statistics. 

In the past two decades, the numerical simulation methods that incorporate cloning algorithms have emerged as alternatives to address this challenge~\cite{Giardina2006,Lecomte2007,PerezEspigares2019}. Although it was initially developed for classical stochastic systems, the cloning algorithm can be integrated with quantum jump trajectories. Recently, this integration has been applied to compute large deviations in the counting statistics~\cite{Carollo2020}. The principle of the wave function cloning algorithm is the correspondence between the time evolution of the tilted quantum master equation and the growth or decay behavior in population dynamics~\cite{Carollo2020}. In the numerical implementation, a population of the quantum systems is simulated simultaneously. During the evolution process, some systems are terminated, whereas other systems are cloned, resulting in a change in the total number of systems. This reflects the nonconservation of the probability distribution governed by the tilted master equations. The concept of the cloning algorithm is general and enables the computation of large deviations for arbitrary dynamical observables. In the current context, it is reasonable to consider applying the cloning algorithm to compute large deviations in the FPT statistics. However, a peculiarity exists in this case: for a current-like counting variable, there are two SCGFs, $\Psi_{\pm}(\nu)$. Consequently, a novel implementation of the wave function cloning algorithm for the FPT statistics is needed. Owing to the involvement of numerous technical details, we present them in Appendix~\ref{CloningAlgorithm}.

\section{Examples}
\label{section6}
We consider two types of quantum systems. The first type consists of single atoms, for which we take typical two-level and three-level systems as examples~\cite{Mandel1995}. The quantum jump trajectories of these systems can be described as a semi-Markov process~\cite{Liu2022}. With respect to resonant driving, the SCGFs in the FPT statistics of the two-level system have been solved using the semi-Markov process method~\cite{Liu2024}. In addition to confirming the consistency between the theories developed in this paper and the previous one, we find that solving the equation of poles~(\ref{equationofpolestiltedquantummasterequation}) is more easily accomplished, even when the nonresonant case is considered. The second type is the interacting few-body system. We take two interacting two-level atoms as an example. In this system, the semi-Markov process description fails.  
 
\begin{figure}
\includegraphics[width=1.\columnwidth]{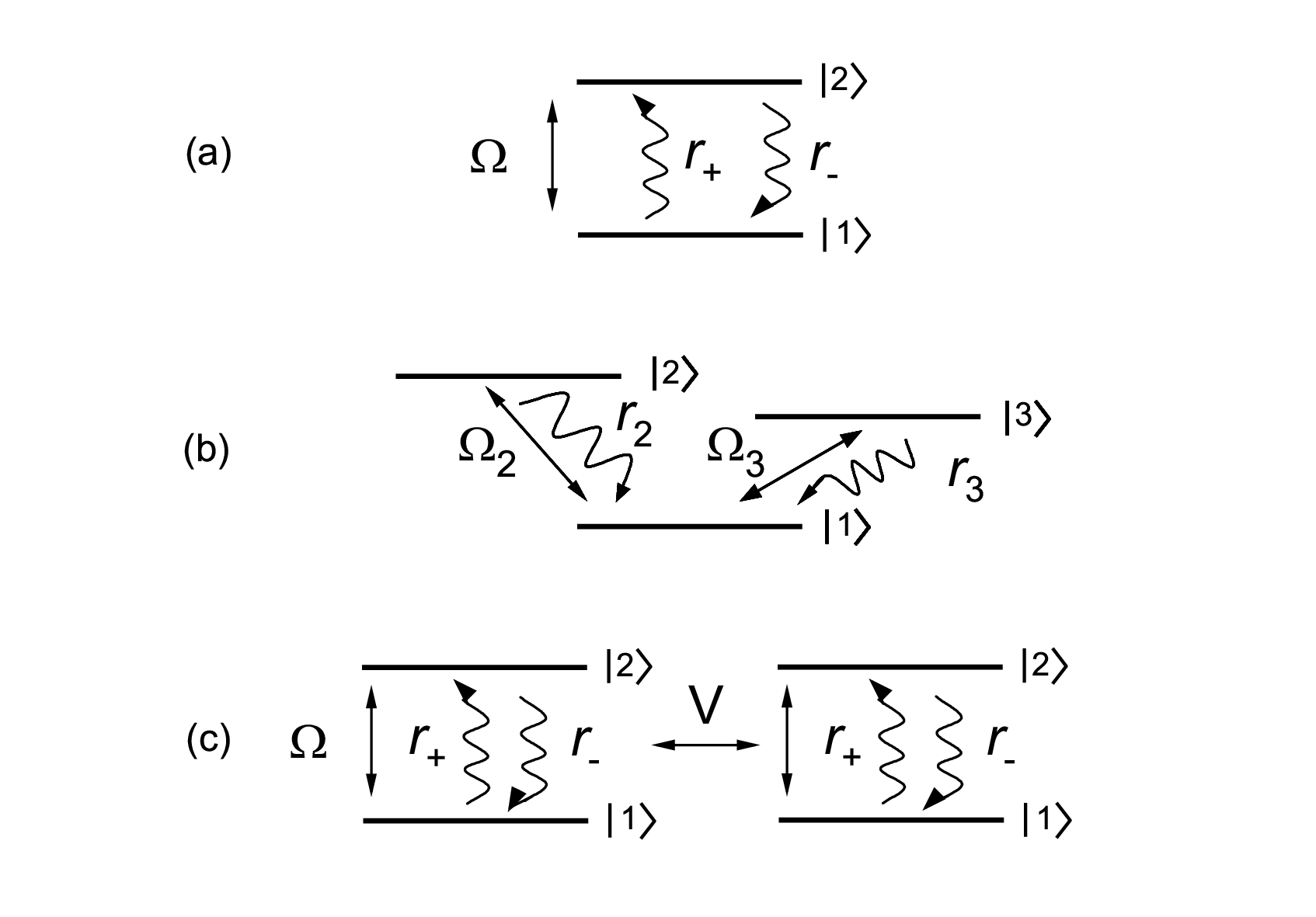}
\caption{Schematics of a two-level system (a), three-level system (b), and two interacting two-level atoms (c). The lines denoted by double arrows indicate driving, whereas the lines denoted by a single arrow represent the jumps occurring within the quantum systems. }
\label{fig1}
\end{figure}
 
\subsection{Two-level system}
\label{section61}
A schematic of the driven two-level quantum system is shown in Fig.~\ref{fig1}(a). The inverse temperature of the surrounding heat bath is $\beta$. In the interaction picture of the rotating frame with the field frequency, the generator of the quantum master equation for the system is expressed as follows:
\begin{eqnarray}
	\label{twolevelquantummasterequation}
	L[\rho]&=&-{\rm i}\left[ H,\rho\right ] + r_-[\sigma_-\rho\sigma_+ -\frac{1}{2}\{\sigma_+\sigma_-, \rho\}   ]\nonumber \\
	&&+r_+[\sigma_+\rho\sigma_- -\frac{1}{2}\{\sigma_-\sigma_+, \rho \}   ],
\end{eqnarray}
where $H=\Delta \sigma_z-\Omega \sigma_x/2$ represents the Hamiltonian, $\Delta$ denotes the detuning of the field, $\Omega$ indicates the Rabi frequency representing the interaction strength between the field and the system, $\sigma_\pm$ represent the raising and lowering Pauli operators, respectively, and $r_\pm$ denote the pumping and damping rates, respectively. These two rates satisfy the detailed balance condition $r_-=r_+\exp{(\beta\omega_0 )}$, where $\omega_0$ is the difference in the energy level.   

Equation~(\ref{twolevelquantummasterequation}) indicates that there are two jump operators, $\sigma_{\pm}$. We consider two cases. In the first case, the weights are, see Eq.~(\ref{countingquantity}), $w_{\pm}=\mp 1$. This counting variable is current-like and is related to the entropy production in the quantum jump trajectories~\cite{Breuer2003,Liu2016a}. By explicitly writing the matrix representation of the tilted generator $\tilde{L}_z$ and performing simple algebraic operators on Eq.~(\ref{equationofpolestiltedquantummasterequation}), we obtain a quadratic equation in the variable $z$: 
\begin{eqnarray}
\label{equationofpolesforTLSentropyproduction}
z^{-2}-2b(\nu) z^{-1}+\frac{ r_+}{r_-}=0,
\end{eqnarray} 
where the coefficient is given by  
\begin{eqnarray}
b(\nu)=\frac{1}{\Omega^2r_-\xi(\nu)}
\left[\xi^4(\nu)+\xi^2(\nu)(4\mu^2-r_-r_++\Delta^2)-\frac{r^2}{4}\Delta^2 \right], 
\end{eqnarray}
with $\xi(\nu)=\nu+r/2$, $r=r_-+r_+$, $16\mu^2=4\Omega^2-\delta^2$, and $\delta=r_--r_+$. According to Eq.~(\ref{SCGFdefinitionFPTtype2positive}), the SCGFs in the FPT statistics are 
\begin{eqnarray}
\label{SCGFofentropycurrent}	\Psi_{\pm}(\nu)=-\ln\left[ b(\nu)\pm\sqrt{b^2(\nu)-\frac{r_+}{r_-} }\right],
\end{eqnarray}
respectively. Because both of these functions are real, $\nu$ has a lower bound. Using Vieda's theorem in the quadratic equation and considering that the two rates satisfy the detailed balance condition, we can directly confirm that these two functions satisfy the fluctuation theorem~\cite{Lebowitz1999}:  
\begin{eqnarray}
\label{fluctuationtheorem}
\Psi_{+}(\nu)=\beta\omega_0 -\Psi_{-}(\nu).  
\end{eqnarray} 
If the detuning $\Delta$ is zero, $\it i.e.$, a resonant field is driving the system, Eq.~(\ref{SCGFofentropycurrent}) yields the same result as Eq.~(37) in Ref.~\cite{Liu2024}. Hence, the detuning does not affect the fluctuation theorem. 

\begin{figure}
	\includegraphics[width=1\columnwidth]{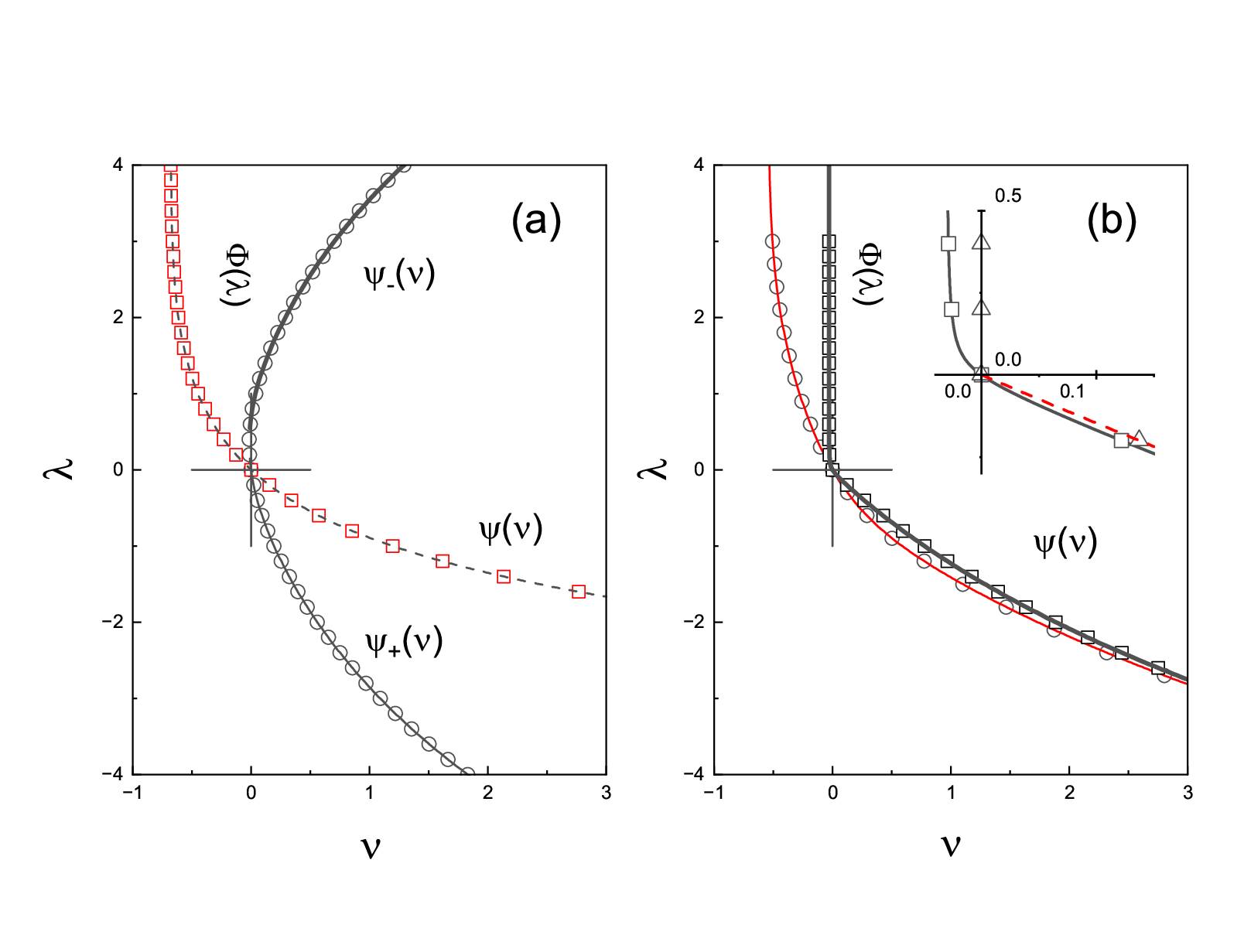}
	\caption{(a) The solid curves (both thick and thin) and dashed curves are the SCGFs $\Psi_{\pm}(\nu)$ and $\Psi(\nu)$ in the FPT statistics given by Eqs.~(\ref{SCGFofentropycurrent}) and~(\ref{SCGFtwolevelsystemdynamicactivity}),  respectively. The open circles and squares represent the SCGFs $\Phi(\lambda)$ in the counting statistics for the corresponding counting variables. The parameters are in arbitrary units: $r_+=0.5$, $r_-=1$, $\Omega=0.8$, and $\Delta=0.2$. (b) The thick and thin curves are the SCGF $\Psi(\nu)$ in the FPT statistics given by Eq.~(\ref{SCGFThreelevelsystemdynamicactivity}). The parameters are in arbitrary units: $\Delta_2=0.3$, $\Delta_3=0.2$, $r_2=4$, $r_3=0$,  $\Omega_2=2$, $\Omega_3=1$ for the thin curve, and $\Omega_3=0.2$ for the thick curve. The open symbols represent the data of the SCGF $\Phi(\lambda)$ for the dynamical activity in the counting statistics for the same sets of parameters. Inset: A magnified view of thick curve and square symbols around the origin $(0,0)$. The dashed curve and open triangle symbols therein are the SCGFs $\Psi(\nu)$ and $\Phi(\lambda)$ for the same parameters except for $\Omega_3=0$, respectively. It should be noted that this curve terminates at the origin. In both panels, the symbol $\Phi(\lambda)$ is placed horizontally, and the large crosses indicate the origins. }
	\label{fig2}
\end{figure}

In the second case, we are interested in the dynamic activity, where both weights $w_{\pm}$ are equal to $1$. Simple algebraic operations on  Eq.~(\ref{equationofpolestiltedquantummasterequation}) yield   
\begin{eqnarray}
\label{equationofpolesforfullcountingvaraible}
	8r_+r_-[\Delta^2+\xi^2(\nu)] z^{-2} + 2r\Omega^2 \xi(\nu) z^{-1}-\left[4\xi^4(\nu)+4\xi^2(\nu)(\Delta^2+4\mu^2)-\delta^2\Delta^2 \right]=0. 
\end{eqnarray}  
This is also a quadratic equation in the variable $z$. According to Eq.~(\ref{SCGFdefinitionFPTtype1}), the SCGF in the FPT statistics is the logarithm of the largest positive real root. Thus, we obtain  
\begin{eqnarray}
\label{SCGFtwolevelsystemdynamicactivity}
\Psi(\nu)=\ln\frac{8r_+r_-[\Delta^2+\xi^2(\nu)]}{\sqrt{4r^2\Omega^4\xi^2(\nu)+16r_+r_-[\Delta^2+\xi^2(\nu)][4\xi^4(\nu)+4\xi^2(\nu)(\Delta^2+4\mu^2)-\delta^2\Delta^2]}-2r\Omega^2\xi(\nu) }. 
\end{eqnarray}
If $\Delta=0$, Eq.~(\ref{SCGFtwolevelsystemdynamicactivity}) yields the same result as Eq.~(42) in the previous paper. In Fig.~\ref{fig2}(a), we numerically illustrate Eqs.~(\ref{SCGFofentropycurrent}) and~(\ref{SCGFtwolevelsystemdynamicactivity}) for a set of parameters. We also calculate the SCGFs $\Phi(\lambda)$ of the corresponding counting variables in the counting statistics: the data are obtained by finding the roots of Eq.~(\ref{equationofpolestiltedquantummasterequation}) with respect to the variable $\nu$ and then applying Eq.~(\ref{SCGFformulacountingstatistics}). The consistency between the curves and open symbols in the figure demonstrates the fourth result in Sec.~(\ref{section4}), {\it i.e.}, the inverse function relationship of the large deviations in the two statistics. According to the Legendre transform~\cite{Touchette2008}, we can infer that for the large FPT, due to the $\nu$-value ($<0$) near zero at the $\Psi_{\pm}$ connecting point ($b(\nu)=\sqrt{r_+/r_-}$) under the given parameters, the distribution of the current-like counting variable has a much longer tail than the dynamic activity's as these counting variables tend to positive infinity, since it's an exponential decay function with the $\nu$-value as rate.

\subsection{Three-level system}
\label{section62}
The other single-atom model is the $V$-type three-level system, which is schematically depicted in Fig.~(\ref{fig1})(b). Assuming its environment is vacuum, in the interaction picture of a rotating frame, the generator of the quantum master equation for the system is expressed as follows:   
\begin{eqnarray}
	\label{threelevelgeneratorofquantummasterequation}
	L[\rho]&=&-{\rm i}[H,\rho]+r_2[\sigma_{12}\rho(t)\sigma_{21} -\frac{1}{2}\{\sigma_{21}\sigma_{12}, \rho(t) \}\nonumber \\
	&&+r_3[\sigma_{13}\rho(t)\sigma_{31} -\frac{1}{2}\{\sigma_{31}\sigma_{13}, \rho(t) \}   ], 
\end{eqnarray}
where $\sigma_{ij}\equiv|i\rangle\langle j|$ for $i,j=1,2,3$. In Eq.~(\ref{threelevelgeneratorofquantummasterequation}), the Hamiltonian of the system is 
\begin{eqnarray}
	\label{HamiltonianthreelevelinV}
	H=\Delta_2\sigma_{22} + \Delta_3 \sigma_{33}+\frac{\Omega_2}{2}(|2\rangle\langle 1|+ +\sigma_{21})+\frac{\Omega_3}{2}(\sigma_{13} +\sigma_{31}),
\end{eqnarray}
where $\Delta_i$ represents the detuning of the $i$-th field, and $\Omega_{i}$ represents the Rabi frequency of the field driving the $|1\rangle$-$|i\rangle$ transition, for $i=2,3$. There are also two jump operators, $\sigma_{12}$ and $\sigma_{13}$. In this case, we are interested in the dynamic activity, $\it i.e.$, the weights $w_\alpha=1$. Once again, by explicitly writing the matrix representation of the tilted generator $L_z$ and performing simple algebraic operators on   Eq.~(\ref{equationofpolestiltedquantummasterequation}), we derive 
\begin{eqnarray}
	\label{equationofpolesforthreelevelSentropyproduction}
	z^{-1}c_1(\nu)-c_0(\nu)=0, 
\end{eqnarray} 
where the two coefficients are 
\begin{eqnarray}
	c_0(\nu)&=&\nu^9+\cdots\\
	c_1(\nu)&=&\left(\frac{\Omega_2^2}{2}r_2+\frac{\Omega_3^2}{2}r_3\right)\nu^6+\cdots,
\end{eqnarray}
respectively. Owing to their length, we do not write them out in full. Thus, the SCGF in the FPT statistics is simply 
\begin{eqnarray}
	\label{SCGFThreelevelsystemdynamicactivity}
	\Psi(\nu)=\ln \frac{c_1(\nu)}{c_0(\nu)}. 
\end{eqnarray}
Owing to the logarithmic function involved, in this model, $\nu$ also has a lower bound.  

We may formally demonstrate the correctness of Eq.~(\ref{SCGFThreelevelsystemdynamicactivity}). Like those of the two-level system, the quantum jump trajectories of the three-level system can be described by the semi-Markov process. Because there is a unique collapsed wave function, $|1\rangle$, the semi-Markov method can be applied to obtain the SCGF in the FPT statistics~\cite{Liu2024}. The result is 
\begin{eqnarray}
\label{SCGFdynamicactivitysemiMarkovmethod}
\Psi(\nu)=\ln\left[ \hat{ p}^{(2)}_{11}(\nu)+{\hat p}^{(3)}_{11}(\nu)\right].  
\end{eqnarray}
Here, $\hat p_{11}^{(i)}$, for $i=2,3$, are the Laplace transforms of the waiting-time distributions $p_{11}^{(i)}(t)$. They represent the possibility that the quantum system starts from $|1\rangle$, successively evolves and eventually collapses back to $|1\rangle$ at time $t$ via the jump operator $\sigma_{1i}$. In a specific case with $r_3=0$ and $\Delta_i=0$, for $i=2,3$, $p_{11}^{(2)}(\nu)$ has been solved exactly~\cite{Liu2023b}. By examining  Eqs.~(\ref{SCGFThreelevelsystemdynamicactivity}) and~(\ref{SCGFdynamicactivitysemiMarkovmethod}), we find that they are indeed consistent with each other.  

In Fig.~\ref{fig2}(b), we illustrate Eq.~(\ref{SCGFThreelevelsystemdynamicactivity}) using three sets of parameters. The data for the SCGF $\Phi(\lambda)$ of the dynamical activity are also shown therein. We can observe that, when both the parameters $\Omega_3$ and $r_3$ are near zero, there is a rapid phase crossover at $\nu=0$ in both large deviation functions; also see the inset in the same panel. This is a smoothed first-order dynamical transition~\cite{Garrahan2010}. According to Eq.~(\ref{SCGFThreelevelsystemdynamicactivity}), the characteristic stems from the fact that the polynomial $c_0(\nu)$ is zero in the region immediately to the left of the origin. On the other hand, consider the case where $\Omega_3$ and $r_3$ are exactly equal to zero. In this situation, the three-level system is known to exhibit an actual dynamical transition in the counting statistics: the SCGF $\Phi(\lambda)$ becomes non-analytical at the origin~\cite{Lesanovsky2013a}; also see the triangle symbols in the inset. Interestingly, under this condition, because the functions $c_0(\nu)$ and $c_1(\nu)$ are exactly zero at $\nu=0$, Eq.~(\ref{equationofpolesforthreelevelSentropyproduction}) degenerates and its solution becomes arbitrary. Consequently, $\Psi(\nu)$ terminates at the origin; see the dashed curve in the inset. From the perspective of the FPT distribution, using the Legendre transform again and carrying out a similar argument in the two-level system, the rapid phase crossover implies that for the large FPT, the distribution has a much longer tail ( an exponential decay function having rate $\nu\approx 0$ and $c_0(\nu)=0$) than other distributions without rapid crossover, such as that of the thin curve in Fig.~(\ref{fig2})(b). Moreover, as the phase transition occurs, because the SCGF $\Psi(\nu)$ terminates at the origin, the tail will tend to infinity, that is, the large deviation principle will break down~\cite{Touchette2008}.

We explain the significance of deriving these analytical formulas for the large deviations in the FPT statistics. Given an arbitrary set of parameters, analytically solving for large deviations in the counting statistics is a challenging task. For example, in seemingly simple two-level and three-level systems, finding the largest real root of Eqs.~(\ref{equationofpolesforTLSentropyproduction}) and~(\ref{equationofpolesforthreelevelSentropyproduction}) with respect to the variable $\nu$ is equivalent to solving the fourth- and ninth-degree polynomials in $\nu$, respectively. This is either too complex or there are no radical solutions to these equations~\cite{Speigel1968}. However, Eqs.~(\ref{SCGFofentropycurrent}),~(\ref{SCGFtwolevelsystemdynamicactivity}), and~(\ref{SCGFThreelevelsystemdynamicactivity}), as well as the bidirectional inversion function relationship between the large deviations in the two statistics, help overcome this difficulty. 

\subsection{Two interacting two-level atoms}
\label{section63}
In the final model, two interacting two-level atoms are uniformly driven by a classical field and are surrounded by a heat bath with an inverse temperature $\beta$; see Fig.~(\ref{fig1})(c). In the interacting picture, the generator of the quantum master equation for this two-body system is~\cite{Lee2012} 
\begin{eqnarray} 
L[\rho]=-{\rm i}\left[ H,\rho\right ] +\sum_{i=1}^2 r_-\left[|1\rangle\langle 2|_i \rho |2\rangle\langle 1|_i -\frac{1}{2}\left\{|2\rangle \langle 2|_i,\rho\right \}\right] \nonumber \\
+\sum_{i=1}^2 r_+\left[|2\rangle\langle 1|_i \rho |1\rangle\langle 2|_i -\frac{1}{2}\left\{|1\rangle \langle 1|_i,\rho\right \}\right]
\end{eqnarray}
where the Hamiltonian is 
\begin{eqnarray}
	H=\sum_{i=1,2}\left[ \Delta |2 \rangle \langle 2|_i +\frac{\Omega}{2} (|1\rangle \langle 2|_i+|2\rangle \langle 1|_i)\right]+ V|2\rangle\langle 2|_1\otimes |2\rangle\langle 2|_2,
\end{eqnarray}
$\Delta$ represents the detuning of the field, $\Omega$ denotes the Rabi frequency of the field, and $V$ represents the interaction strength between the two atoms. There are four types of jump operators,  $|1\rangle \langle 2|_i$, and $|2\rangle \langle 1|_i$, for $i=1,2$. Because these operators act only on individual atoms and are local, the semi-Markov process cannot characterize the quantum jump trajectories of this system. A brief explanation is provided in Appendix~\ref{failuresemiMarkovdescription}.  

\begin{figure}
\includegraphics[width=1\columnwidth]{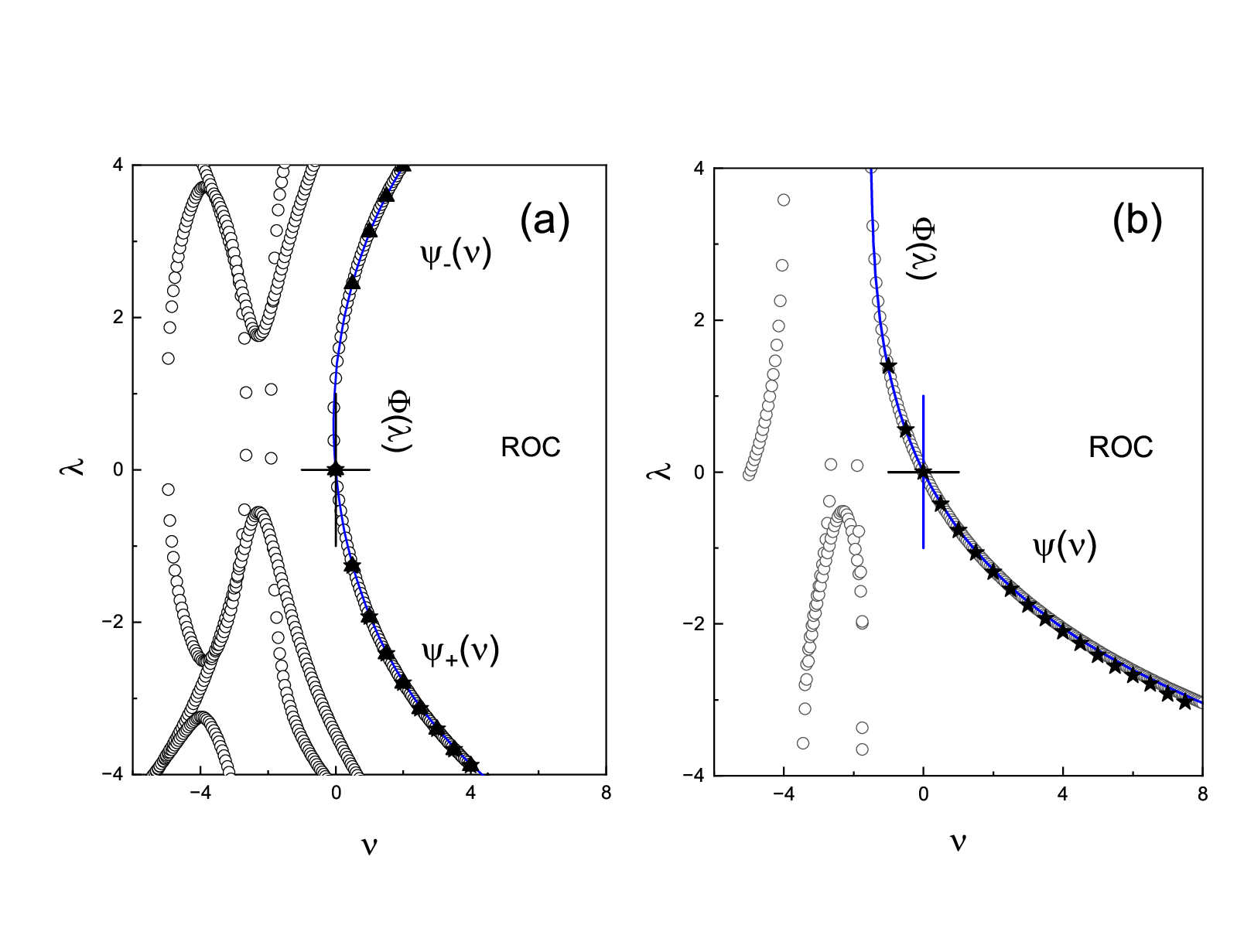}
\caption{For the two interacting two-level atoms, the open circles in the real $\nu$-$\lambda$ plane represent the real solutions of the equation of poles~(\ref{equationofpolestiltedquantummasterequation}), where $\nu$ represents free parameter and $z(\nu)$ represents the real solutions or roots. We obtain these data at discrete $\nu$-values and do not connect them with curves. Nevertheless, we can still easily identify the continuous boundaries of the regions of convergence (ROCs). Both boundaries lie on the rightmost side of the plane. They pass through the point $(0,0)$ (denoted by large crosses), and extend toward the right-hand side. Thus, $\lambda(\nu)$ on these boundaries precisely corresponds to the SCGFs $\Psi_{\pm}(\nu)$ or $\Psi(\nu)$ in the FPT statistics. The solid curves represent the SCGFs in the counting statistics, $\Phi(\lambda)$. These are obtained by finding the roots of the same Eq.~(\ref{equationofpolestiltedquantummasterequation}) with respect to the variable $\nu$ and then applying Eq.~(\ref{SCGFformulacountingstatistics}). In this case, the free parameter is $z=\exp(\lambda)$. The solid triangle and asterisk symbols denote the SCGF data obtained from simulations built on the wave function cloning algorithm. The fixed small time step for the quantum trajectory simulation is $dt=0.01$, and $10000$ initial copies of the quantum system are used for the population dynamics. (a) Current-like counting variable. (b) Simple counting variable. Their respective weights are described in the main text. The parameters are in arbitrary units: $\Delta=0.1$, $\Omega=1$, $V=5$, $r_-=2.0$, and $r_+=0.6$.   }
\label{fig3}
\end{figure}

Owing to the presence of a pair of two-level atoms, the matrix representation of the tilted generator $L_z$ is a $16\times16$ matrix. First, we focus on a current-like counting variable with weights $w_{|1\rangle\langle 2|_{i}}=1$ and $w_{|2\rangle\langle 1|_{i}}=-1$, for $i=1,2$, respectively. Substituting these weights into the equation of poles, Eq.~(\ref{equationofpolestiltedquantummasterequation}), we obtain a tenth-degree polynomial in the variable $z$, which is also a sixteenth-degree polynomial in the variable $\nu$. To find the roots of this equation with respect to $z$, a numerical algorithm is needed. In Fig.~(\ref{fig3})(a), with a specific set of physical parameters, we plot all $\lambda(\nu)$, $\it i.e.$, the logarithms of the positive real roots $z(\nu)$, where $\nu$ is considered a free parameter. According to the first and third results in Sec.~(\ref{section4}), we can easily identify the boundary of the region of convergence. Consequently, the SCGFs $\Psi_{\pm}(\nu)$ in the FPT statistics are equal to $\lambda(\nu)$ on the boundary curve. To independently confirm these data, we perform a simulation using the wave function cloning algorithm; see the solid triangle symbols and asterisk symbols in Fig.~(\ref{fig3})(a)~\footnote{	The data are not publicly available. The data are available from the authors upon reasonable request.}. We observe that the data obtained by the two methods are in agreement. In addition, the fluctuation theorem~(\ref{fluctuationtheorem}) is numerically validated to hold in this scenario. Second, we consider the simple counting variable with weights $w_{|1\rangle\langle 2|_{i}}=1$ and $w_{|2\rangle\langle 1|_{i}}=0$, for $i=1,2$, respectively. In this case, the equation of poles is a fifth-degreee polynomial in the variable $z$. We perform a similar analysis and present the data in Fig.~(\ref{fig3})(b). We notice that the boundary of the region of convergence for the simple counting variable extends in the $\lambda$-direction. This is quite different from the arch-shaped boundary in the case of the current-like counting variable. Furthermore, the SCGF computed either by taking the logarithm of the $z(\nu)$-value on the boundary of the region of convergence or using  Eq.~(\ref{SCGFdefinitionFPTtype1}) is in agreement with the data obtained from the simulation method. 

\section{Conclusion}
\label{section7}
Inspired by the inverse function relationship between SCGFs in counting statistics and those in FPT statistics for open quantum systems, this paper investigates alternative approaches for computing the large deviations in FPT statistics. We develop two methods. The first method solves the equation of poles with respect to the $z$-transform parameter to identify the boundary of the region of convergence. The SCGFs in the FPT statistics are then simply the logarithm of the $z$-value on this boundary. The second method is a simulation-based approach built upon the wave function cloning algorithm. When the Hilbert space increases exponentially with increasing degrees of freedom, the simulation method becomes indispensable. Our motivation for developing the two methods independently of the inverse function relationship stems from the self-completeness of the theory. Indeed, we note that the FPT statistics is related to, rather than equivalent to, the counting statistics. In particular, computing the large deviations in the latter statistics is far from trivial. 

Let us conclude this paper by briefly comparing the equation of poles of the semi-Markov process with the equation of poles derived in this paper. In open quantum systems that can be characterized by the semi-Markov process, these two equations clearly yield completely consistent large deviation results. The analytical solutions for the two-level and three-level systems illustrate this aspect. When examining the matrices in computation, the size of the matrix in the former case is equal to the square of the number of collapsed quantum wave functions, whereas in the latter case it is the fourth power of the dimension of the Hilbert space. Consequently, the matrix involved in the equation of poles of the semi-Markov process is substantially smaller than that of the equation of poles presented in this paper. However, the cost of this is that the semi-Markov process method requires solving the dynamics of the quantum system to obtain the probability-based waiting-time distributions. Even for a two-level system, if it involves arbitrary parameters, this step is very cumbersome to accomplish through manual calculation. The advantage of the equation of poles derived in this paper is that it does not involve dynamic problems.\\ 
\\
{\noindent\it Acknowledgments}
This work was supported by the National Natural Science Foundation of China under Grant No. 12075016 and No. 12505048.

\appendix 
\section{Operator $G$ and transition rates in Eq.~(\ref{tiltedLiouvillmasterequation})}
\label{detailsintiltedLiouvillemasterequation}
The action of the operator $G$ on a wave function $\psi$ is expressed as follows: 
\begin{eqnarray}
\label{Goperator}
G[\psi]=\left(H_{\rm eff}  + \frac{\text{i}}{2}\sum_{\alpha=1}^M r_\alpha \parallel A_\alpha\psi\parallel ^2\right ) \psi,
\end{eqnarray}
where 
\begin{eqnarray}
\label{effectiveHamiltonian}
H_{\rm eff}\equiv H-({\text{i}}/{2})\sum_{\alpha=1}^M r_\alpha {A_\alpha}^\dag A_\alpha 
\end{eqnarray}
represents the non-Hermitian Hamiltonian. In the second integral of Eq.~(\ref{tiltedLiouvillmasterequation}),  the tilted transition rate is defined as follows: 
\begin{eqnarray}
\label{rateauxiliarysystem}
W_z[\phi|\psi]=\sum_{\alpha=1}^M \text{z}^{-\omega_\alpha} k_\alpha[\phi]\delta\left[\frac{A_\alpha \phi }{\parallel A_\alpha \phi\parallel }-\psi\right], 
\end{eqnarray}
where $\delta[\cdots]$ denotes the Dirac functional and
\begin{eqnarray} 
	k_\alpha[\phi]= r_\alpha \parallel A_\alpha \phi\parallel ^2
\end{eqnarray}
is the jump rate of the jump operator $A_\alpha$. The transition rate $W[\cdots]$ in the same integral is equal to Eq.~(\ref{rateauxiliarysystem}) with $z=1$.  

Equation~(\ref{tiltedLiouvillmasterequation}) is derived by applying the Feynman-Kac formula in the Hilbert space and a distribution functional transformation of the quantum jump trajectory. Owing to its abstract nature, this equation is rarely applied in practical evaluations. However, an ``averaged'' operator,  
\begin{eqnarray}
	\label{tiltedreduceddensitymatrix}
	\tilde {\rho}(z|t)=\int D\psi D\psi^*\bar {\cal P} [\psi,z|t] |\psi\rangle \langle \psi|, 
\end{eqnarray}
precisely satisfies the tilted quantum master equation:
\begin{eqnarray}
	\label{tiltedquantummasterequation}
\partial_t \tilde{\rho}=\tilde {L}_z[\tilde{\rho}],
\end{eqnarray}
where the generator $\tilde{L}_z$ is defined in Eq.~(\ref{eigenoperatoroftiltedquantumasterequation}). Because the Liouville master equation is the unraveling of the quantum master equation in the Hilbert space~\cite{Breuer1995a}, Eq.~(\ref{tiltedLiouvillmasterequation}) is also the unraveling of the tilted quantum master equation in the same space. Eqs.~(\ref{tiltedreduceddensitymatrix}) and~(\ref{tiltedquantummasterequation}) form the basis of the equation of poles in the averaged dynamics; see the discussion in Sec.~(\ref{section41}). More details on the tilted Liouville master equation can be found in a previous paper~\cite{Liu2021a}. 

\section{Wave function cloning algorithm}\label{CloningAlgorithm}

\subsection{Simulating quantum jump trajectories}
For the convenience of the reader, because the wave function cloning algorithm for computing the large deviations integrates the simulation of the quantum jump trajectory, we briefly review how to simulate quantum jump trajectories~\cite{Breuer2002}. The wave function of an individual quantum system undergoes a deterministic evolution interrupted by random jumps. Specifically, a single quantum jump trajectory proceeds as follows:
\begin{itemize}
\item[(1)] Propagate the wave function $\psi(t)$ forward for a small time step $dt$ using the effective non-Hermitian Hamiltonian $H_{\rm eff}$ in  Eq.~(\ref{effectiveHamiltonian}).
\item[(2)] Calculate the probability
\begin{eqnarray}
    p=r_\alpha dt\sum_{\alpha}\langle {\psi(t+{ d}t)|A_{\alpha}^{\dagger}A_{\alpha}|\psi(t+{d}t)\rangle} \text{.} \nonumber
\end{eqnarray}
\item[(3)] Draw a random number $r\in [0,1)$. If $r<p$, a quantum jump occurs and the jump type is determined by the relative probability of $r_\alpha\langle{\psi(t+{ d}t)|A_{\alpha}^{\dagger}A_{\alpha}|\psi(t+{d}t)}\rangle$. Then $|\psi\rangle\gets A_\alpha|{\psi}\rangle$, where $A_\alpha$ represents the jump operator that triggers the quantum jump. If $r>p$, no quantum jump occurs. 
\item[(4)] Normalize the wave function so that $\langle{\psi|\psi}\rangle=1$ and return step (1) to continue.
\end{itemize}
The time step ${\rm d}t$ should be small enough to ensure that quantum jumps occur in quasi-continuous steps.

\subsection{Cloning algorithm for counting statistics} 
The wave function cloning algorithm for computing large deviations in the FPT statistics is a variant of that used in the counting statistics~\cite{Carollo2020}. It is beneficial to first present the latter algorithm. In this context, population dynamics is simulated, and the additional steps involve identifying the cloning factor, making clones according to this factor, and maintaining the number of systems. Initially, we prepare $N_0$ copies of the system. In addition to state information, each system copy has a time variable $t$ that is initially zero. The cloning algorithm consists of the following steps:
\begin{itemize}
\item[(1)] For each system ${\cal S}$, a quantum trajectory is propagated forward in time by step $dt$. Consequently, its time variable is updated so that $t\gets t+dt$. If a quantum jump of type $\alpha$ occurs, the cloning factor is $Y = e^{-\lambda w_{\alpha}}$, where $w_{\alpha}$ represents the corresponding weight. If no quantum jump occurs, then $Y=1$.
\item[(2)] The system ${\cal S}$ is either terminated or cloned on the basis of the cloning factor:
\begin{itemize}
\item[\textbullet] calculates $y=[Y+\epsilon]$, where $[\cdot]$ denotes the operator of taking the integer part and $\epsilon$ is uniformly distributed in the range $(0,1)$;
\item[\textbullet] if $y=0$, the system ${\cal S}$ is terminated;
\item[\textbullet] if $y>1$, we make $y-1$ new clones of ${\cal S}$, all of which are in the updated state.
\end{itemize}
\end{itemize}
The above procedure may result in either exponential growth or complete decay of the number of systems. Therefore, we add a third step to keep the number of systems at $N_0$:
\begin{itemize}
\item[(3)] After each time step $dt$, we have $N$ copies of the systems. If $N>N_0$, we randomly terminate the $N-N_0$ systems. If $N<N_0$, we make $N_0-N$ random clones from the remaining copies. We record the value $X=N/N_0$. Then, the procedure continues by looping over the above steps.
\end{itemize}
For an evolution with a long interval $t$, we record many $X$ values: $\{X_1,X_2,\cdots\}$. The SCGF is then computed from the long-time behavior of the product of all these values as follows:
\begin{align}
\Phi(\lambda)=\lim_{t\to\infty}\frac{1}{t}\ln\left\langle{\rm e}^{-\lambda n(t)}\right\rangle\approx\frac{1}{t}\ln\left(X_1X_2\cdots\right) \text{.}
\end{align}
Because all quantum jump trajectories propagate forward in time in discrete steps $dt$, the cloning approach based on them is also discrete with respect time.

\subsection{Cloning algorithm for FPT statistics}
We can now present the wave function cloning algorithm in the FPT statistics. In this case, the algorithm is slightly more complex. The main reason is that, in FPT statistics, the roles of time and the counting variable are interchanged, and unlike time, the change in the counting variable is stochastic.

If the counting variable is current-like, its value can increase or decrease. Initially, we prepare identical copies of the system, $N_0$. Unlike in the counting statistics, each system copy is associated with a variable $\tilde{n}$ that is also initially zero. The procedure of the cloning algorithm for computing the SCGF in the FPT statistics is presented as follows:
\begin{itemize}
	\item[(1)] For each system ${\cal S}$, propagate its wave function in time via step $dt$. If a quantum jump of type $\alpha$ occurs, update the variable $\tilde{n}$ so that $\tilde{n}\gets\tilde{n}+w_{\alpha}$, where $w_{\alpha}$ represents the corresponding weight of the jump. The cloning factor is $Y={e}^{-\nu{d}t}$, where $\nu$ represents the conjugate parameter of time.
	\item[(2)] Terminate or clone the system ${\cal S}$ according to the cloning factor. This step is the same as that in the counting statistics case.
	\item[(3)] Continue the procedure by looping over steps (1) and (2) until the stochastic variable $\tilde{n}$ of all system copies reaches the predetermined threshold $n$. For each system in this loop, its evolution is frozen once the system reaches this threshold.~\footnote{In numerical simulation, this can be achieved by placing all system copies in a dynamic (length-variable) array. If a copy is terminated, we delete the corresponding array element. If new clones are created, they are added at the end of the array. The loop over steps (1) and (2) is implemented by looping over the array until the threshold value $n$ is reached by all copies (array elements).}.
	\item[(4)] When the above loop is terminated, the stochastic variable $\tilde{n}$ of all system copies takes the same predetermined threshold value $n$, and at this point, we have $N$ copies. Maintain the number of systems at $N_0$ using the same method as presented for the counting statistics case. Record the value $X=N/N_0$. Then update the threshold value to a new value $n\gets n+\Delta n$, where $\Delta n$ represents the fixed step for the threshold value. Continues the above procedure by looping over all the above steps.
\end{itemize}
The population dynamics proceeds by looping over the steps multiple times until a final threshold value $n$ is reached. In this process, many $X$ values are recorded: $\{X_1,X_2,\cdots\}$. The SCGF in the FPT statistics is then calculated from the long-time behavior of the product of these values: 
\begin{align}	\Psi(\nu)=\lim_{n\to\infty}\frac{1}{n}\ln\left\langle{\rm e}^{-\nu t(n)}\right\rangle\approx\frac{1}{n}\ln\left(X_1X_2\cdots\right) \text{.} \label{SCGF-FPTS}
\end{align}
There are two loops in the above procedure. The inner loop is guided by incrementing the threshold value $n\gets n+\Delta n$. In this loop, the time it takes is stochastic, manifesting the stochastic nature of FPT $t(n)$. In addition, the variable $\tilde{n}$ is used to emphasize its status as a stochastic variable in numerical simulations, while the threshold value $n$ serves as a parameter guiding the inner loop. Both the FPT $t(n)$ and $\tilde{n}$ derive their stochastic nature from the same origin. Since the difference in the numerical implementations of the algorithm for the FPT statistics and that for the counting statistics lies essentially in the swapped roles of time and the counting variable, the two cloning algorithms exhibits approximately the same convergence speed.

In this paper, we consider only the case in which the weight of any jump can be $0$ and $\pm 1$. Consequently, the fixed step for increasing the threshold $n$ can be $\Delta n=\pm 1$. If the weight is only $0$ or $+1$, the variable $n$ is simple and increases monotonically; thus $\Delta n=1$. In this case, the SCGF in the FPT statistics has only one branch and can be computed from Eq.~(\ref{SCGF-FPTS}), where the final threshold value $n$ is a large and positive. If the weight is $0$ and both $\pm 1$, the variable $n$ is current-like, that is, it can increase ($\Delta n=1$) or decrease ($\Delta n=-1$). In this case, there are two branches of the SCGF in the FPT: $\Psi_{\pm}(\nu)$ when the final threshold value $n$ in Eq.~(\ref{SCGF-FPTS}) is a large and positive or negative. Notably, $\tilde{n}$ represents a stochastic variable, and in numerical simulations over a sufficiently long time, it can have any integer value. This ensures that the inner loop of the procedure can be guided by $n\gets n\pm 1$ in both the positive and negative directions.

\section{Failure of semi-Markov process description on two interacting two-level atoms}
\label{failuresemiMarkovdescription}
At any time $t$, the wave function of the quantum jump trajectory can always be expanded as  
\begin{eqnarray}
	\label{expandedwavefunctionoftwointeractingtwolevelatoms}
	\psi(t)=\sum_{i,j=1}^2 c_{ij}(t) |ij\rangle,
\end{eqnarray} 
Assume a quantum jump or collapse occurs at this moment. The wave function after collapse is proportional to the jump operator acting on Eq.~(\ref{expandedwavefunctionoftwointeractingtwolevelatoms}). For example, if the jump operator is $|1\rangle \langle 2|_1$, the collapsed wave function is proportional to 
\begin{eqnarray}
	|1\rangle\langle 2|_1\psi(t)= c_{21}(t) |11\rangle + c_{22}(t) |12\rangle. 
\end{eqnarray}
Evidently, this wave function is not discretely distributed. In particular, the collapsed quantum system retains its previous ``memory'' through the time-dependent coefficients $c_{21}$ and $c_{22}$. That is, the possibility of the next jump is affected by earlier quantum jumps. Therefore, describing the quantum jump trajectories of the quantum system consisting of two interacting two-level atoms as a semi-Markov process is inappropriate.   
 

\end{document}